RESEARCH ARTICLE

# A Telecare System for Use in Traditional Persian Medicine

Vahid R. Nafisi[1,*] and Roshanak Ghods[2]

[1]*Department of E&IT, Biomedical Engineering Group, Iranian Research Organization for Science and Technology, Tehran, Iran*
[2]*Research Institute for Islamic and Complementary Medicine, School of Persian Medicine, Iran University of Medical Sciences, Tehran, Iran*

**Abstract:**

*Background:*

In Persian Medicine (PM), measuring the wrist temperature/humidity and pulse is one of the main methods for determining a person's health status and temperament. An important problem is the dependence of the diagnosis on the physician's interpretation of the above-mentioned criteria. Perhaps this is one reason why this method has yet to be combined with modern medical methods. Also, sometimes there is a need to use PM to diagnose patients remotely, especially during a pandemic. This brings up the question of how to implement PM into a telecare system. This study addresses these concerns and outlines a system for measuring pulse signals and temperament detection based on PM.

*Methods:*

A system was designed and clinically implemented based on PM that uses data from recorded thermal distribution, a temperament questionnaire, and a customized device that logs the pulse waves on the wrist. This system was used for patient care *via* telecare.

*Results:*

The temperaments of 34 participants were assessed by a PM specialist using the standardized Mojahedi Mizaj Questionnaire (MMQ). Thermal images of the wrist in the supine position (named *Malmas* in PM), the back of the hand, and the entire face were also recorded under the supervision of the physician. Also, the wrist pulse waves were evaluated by a customized pulse measurement device. Finally, the collected data could be sent to a physician via a telecare system for further interpretation and prescription of medications.

*Conclusion:*

This preliminary study focused on the implementation of a combinational hardware-software system for patient assessment based on PM. It appears that the design and construction of a customized device that can measure the pulse waves, and some other criteria, according to PM, is possible and can decrease the dependency of the diagnostic to PM specialists. Thus, it can be incorporated into a telemedicine system.

**Keywords:** Persian medicine, Temperament, Pulse taking, Pulse signal, Thermography, Telecare.

## 1. INTRODUCTION

The WHO recognizes that traditional, complementary, and alternative medicine has many benefits for the potential prevention, control, and treatment of many diseases [1, 2]. Recently, traditional medicines have been considered as an alternative approach to treating COVID-19 [3, 4].

Nowadays, recognizing the physical and psychological differences between people has led to the formation of new branches of modern medicine, such as "personalized medicine" [5]. Despite the fact that this attitude is a new approach in modern medicine, traditional medicine typically pays special attention to these personal differences, and most of their prescriptions are based on a patient's temperament [6].

In Persian Medicine (PM), temperament is a key concept in defining human health and disease; in many diseases, certain changes occur in a person's temperament that can be differentiated according to some principles. Accurate recognition of the patient's dystemperament and determining any deviation from his/her appropriate temperament help physicians to diagnose and treat the disease in PM. If we can accurately categorize patients based on their temperament and treat them with a medicinal plant having the opposite temperament to the disease, then we can increase the patients'

* Address correspondence to this author at the Department of E&IT, Biomedical Engineering Group, Iranian Research Organization for Science and Technology, Tehran, Iran; Tel: +98 21 56276311, E-mail: vr_nafisi@irost.org

satisfaction and reduce the course of the disease [7].

Temperament assessment is usually based on qualitative criteria. In PM, ten items such as wrist temperature (Touch/*Malmas*), the state of the muscle and fat, hair condition, skin color, physique, sleep and wakefulness, physical functions, quality of body waste matter, and the psychic functions are assessed to determine a patient's temperament [8]. Also, pulse measurement, especially the pulse of the radial artery in the wrist, is very important and forms one of the important bases of diagnosis in traditional medicine [9 - 11]. Several of these factors depend on the level of expertise of the physician and environmental conditions such as the temperature of the doctor's hand during the examination, which affects the accuracy and reproducibility of temperament assessment [12].

In PM, the lack of a reliable device that can measure and analyze temperature, humidity, and pulse in accordance with the principles of Persian medicine is strongly felt. This issue can be an obstacle to the development of PM. So far, the need for physical contact between the physician and patient has been a drawback in creating a virtual and remote care system. Today, it is possible to provide the highest quality diagnosis and treatment services even in remote areas via the internet. For this purpose, medical devices with remote connection ability have been provided. The patient can easily be evaluated and monitored with minimal risk of transmitting infectious diseases. Telecare is the best solution for patient care during infectious outbreaks such as SARS, influenza, and COVID-19.

Therefore, the purpose of this study is to investigate the possibility of implementing a telecare system based on PM that uses a thermal camera for temperature/humidity measurement and a customized device to measure the pulse characteristics in the wrist.

## 2. MATERIALS AND METHODS

### 2.1. Participants

This pilot study, based on the medical ethics committee approval, was conducted at the traditional healthcare center of Iran University of Medical Sciences in the winter of 2020 involving 34 volunteers with different temperaments (warm/cold and dry/wet). The volunteers met the inclusion criterion, which was no history of the disease. After obtaining written consent from the volunteers, a PM specialist assessed their temperament according to the standardized Mojahedi Mizaj Questionnaire (MMQ). This questionnaire was standardized in 2014 for Persian medicine by Mojahedi *et al*. [13]. Also, the thermal imaging of their wrist, back of the hand, and the entire face was recorded under the supervision of the PM specialist. In addition, the wrist pulse of each patient was recorded at 7 points on the wrist using the customized pulse-taking device made by this research group. The exclusion criterion included poor signal/image quality.

### 2.2. Tools

In order to classify the patients based on the temperature and humidity of the patient's skin (according to the principles of PM), thermal images of the desired areas were recorded and transferred to the local computer. The thermal camera used in this study was the ULIRVISION-T2 camera, which can receive radiation in a range of 7.5 to 14 micrometers. The imaging was performed in a room under controlled ambient conditions ($23\pm1^{\circ}$C, 25% RH). There was no heat source near the subject. All participants were asked to remove all jewelry. The individual was then asked to sit on a chair, and his/her hand was positioned on a white surface; then, after five minutes (for thermal equilibrium), the imaging was performed. After that, the patient was asked to wear the pulse-taking device and her/his wrist pulses were taken (Fig. **1**).

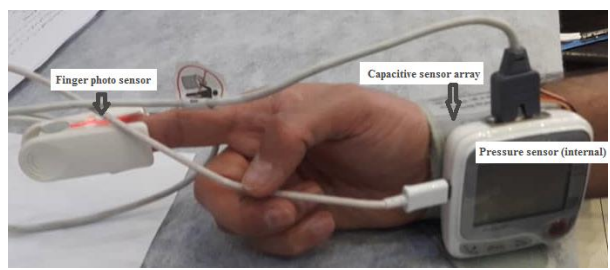

**Fig (1).** Pulse-taking device.

The pulse-taking device has two general parts:

- Measurement hardware, which is equivalent to the sense of touch and pressure applied to a patient's wrist by a traditional medicine doctor. The hardware components of the device include

● An inflation pump for creating different pressures on the pulse position,

● Seven capacitive sensors that measure the passage of stroke volume from different parts of the wrist,

● An optical sensor that measures the intensity of light passing through the patient's finger,

● A pressure sensor that measures and controls the pressure applied to the patient's wrist.

- Pulse signal recording/pre-processing software (the software for the interpretation and diagnosis of diseases and temperaments implemented on the remote computer).

These data (pulse signals and thermal images, together with MMQ) were transferred to the remote system. The main structure of the whole system is shown in Fig. (**2**).

### 2.3. Signal and Image Analysis

After transferring the signals and the images into the remote computer, the signal and image analyses were done in the MATLAB toolbox (Math Works, R2017b).

Various kinds of information concerning the temperature fluctuations and spatial temperature distribution in each area could be obtained. Finally, two sets of features from these areas were extracted:

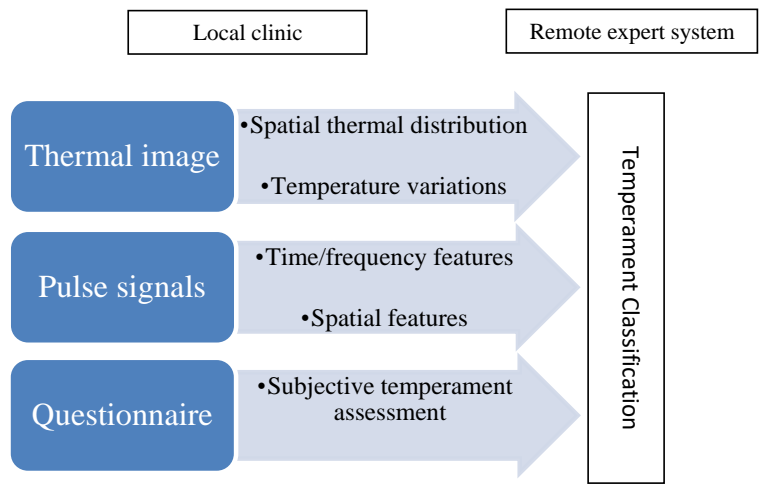

**Fig. (2).** System diagram and flowchart of activities in the telecare system.

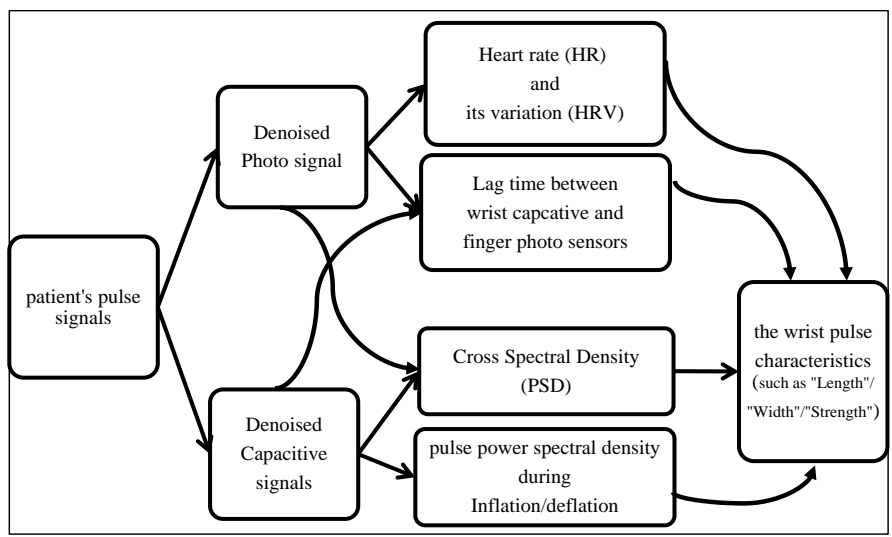

**Fig. (3).** The features extracted from wrist capacitive sensors and finger photo sensor.

1- The features related to the temperature and its fluctuations in the desired areas (13 temperature characteristics) which could be used to discriminate between warm/cold temperaments [14].

2- The features related to the temperature distribution (12 features) at the surface of the areas depending on the characteristics of the tissue that could be used to discriminate between dry/wet temperaments [15].

In addition, rich information could be extracted from the pulse signals (the capacitive sensors on the wrist and optical sensor on the finger), such as

1- Heart rate

2- Pulse strength in each of the measurement positions

3- Temporal changes in pulse power

4- Changes in pulse strength when the wrist is pressurized

5- The lag time between capacitive and optical sensors

Fig. (**3**) shows the features extracted from wrist capacitive sensors and finger photo sensor; the features mentioned in the figure are described as follows:

- Heart rate was calculated from peak-to-peak area of the finger photo sensor. In addition, the power spectra of this signal was mentioned as a representative of the strength of patient's heart activity.
- In PM, the strength of the pulse at each point on the wrist is a very important diagnostic criterion. For this, the power spectral cross-correlation of each capacitive sensor and finger photo sensor (as a reference signal) was used. The cross-spectral density (CSD) is one of the several features used to compare the signals. It shows how correlated two signals are with reference to one another:

$$S_{x,y} = \int_{-\infty}^{+\infty} R_{x,y}(\tau) e^{-i2\pi f\tau} \, d\tau \qquad (1)$$

$S_{x,y}$ and $R_{x,y}$ represent cross-spectral density and cross-correlation between signal x, y (capacitive and photo sensor), respectively.

- Lag time was calculated as the cross-correlation between each capacitive sensor and photo sensor:

$$\text{Lag time} = \tau \mid R_{xy} = max \quad (2)$$

## 3. RESULTS

The demographic information of the 34 participants is listed in Table **1**. In order to achieve a 90% confidence interval in this pilot study, about 65 participants should have been examined [16], but the final sample size was limited to 34 participants (with about 75% confidence interval) due to the COVID-19 epidemic in Iran.

**Table 1. Demographics of participants.**

| Age (Years) | 37.11 ± 7 |
|---|---|
| Gender | 21 females and 13 males |
| Temperament state (based on MMQ) | Warm: 15<br>Moderate: 17<br>Cold: 2 |

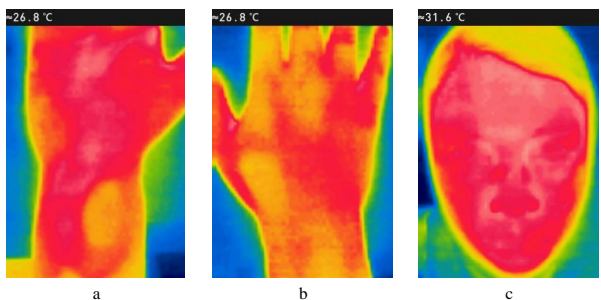

**Fig. (4).** Thermal images of (**a**) wrist; (**b**) back of the hand; (**c**) face.

In this study, we used a combination of capacitive sensors placed on the wrist as an array of sensors to record the wrist pulse signals and their strength (Fig. **5**), the picture on the right). The optical sensor also recorded the pulse signal from the patient's finger. The specification of the collected pulse signals is shown in Table **2**.

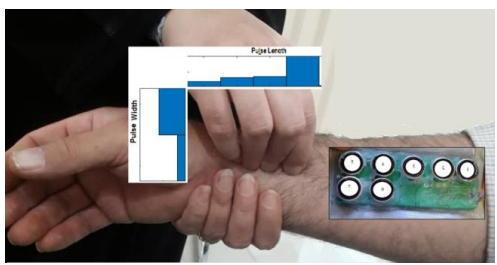

**Fig. (5).** Wrist pulse "length" and "width".

Based on the pulse signal strength in each of the capacitive sensors, a spatial image of the pulse wave propagation along the patient's wrist can be obtained (Fig. **5**), which is known in PM as the pulse length and width.

**Table 2. Data acquisition specification.**

| Acquisition Time (minute) | 1 – 1.5<br>(for each person) |
|---|---|
| Pressure (mmHg) | 0 – 180 |
| Capacitive and photosensors | 200 samples/s<br>Low pass filter: 20 Hz |

In this figure, the horizontal axis of each plot is the capacitive sensor number and the vertical axis is the signal strength in the corresponding sensor (based on Equation 1).

The changes in the wrist pulse signal strength when pressure is applied to the wrist provide important information regarding the strength of the patient's cardiovascular system. When pressure is applied on the wrist, the wrist pulse signal reaction is related to the strength of the cardiovascular system. In this study, in terms of pressure on the wrist, three phases can be considered: no-pressure phase (phase 1), inflation phase (phase 2), and deflation phase (phase 3). Equation 1 was calculated for each capacitive sensor during these phases.

It can be said that when the capacitive sensors receive the pulse signal at higher pressures, it indicates higher strength of the cardiovascular system in the patient, and this is one of the diagnostic indicators in PM. Pressure is applied on the wrist by inflating the cuff around the wrist (Fig. **6**).

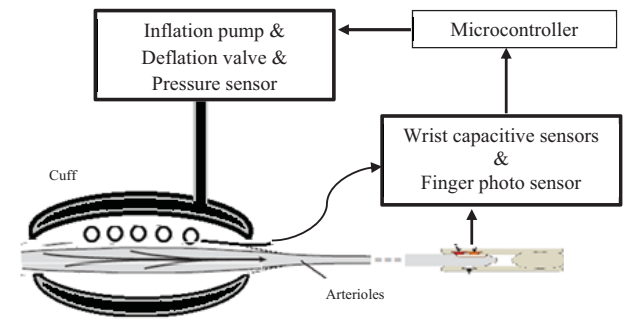

**Fig. (6).** Inflation/deflation devices used for pressure manipulation on the wrist.

Fig. (**7a-c**) shows the changes in the received signals in the capacitive and optical sensors at different pressure phases. (Fig. **7b**) shows that some capacitive sensors received stronger signals after applying the pressure, which meant a deeper pulse in the language of PM.

Figs. (**8**) and (**9**) show the spatial propagation status of the pulse on the wrist in each of the inflation and deflation phases, respectively. Each bar in these figures has been calculated by Equation 1 for capacitance sensor $C_i$ in applied pressure P:

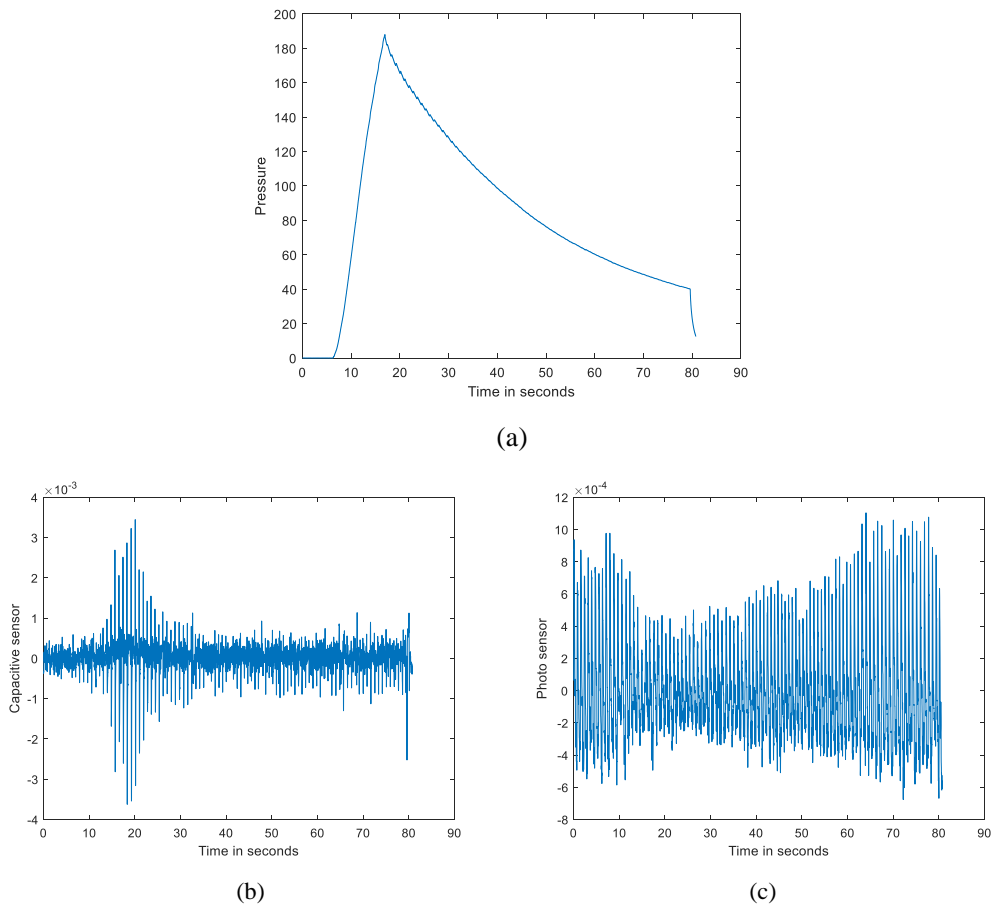

**Figure (7).** Sensors variation during pressure changes. (**a**) pressure phases (**b**) capacitive sensors (**c**) photo sensor.

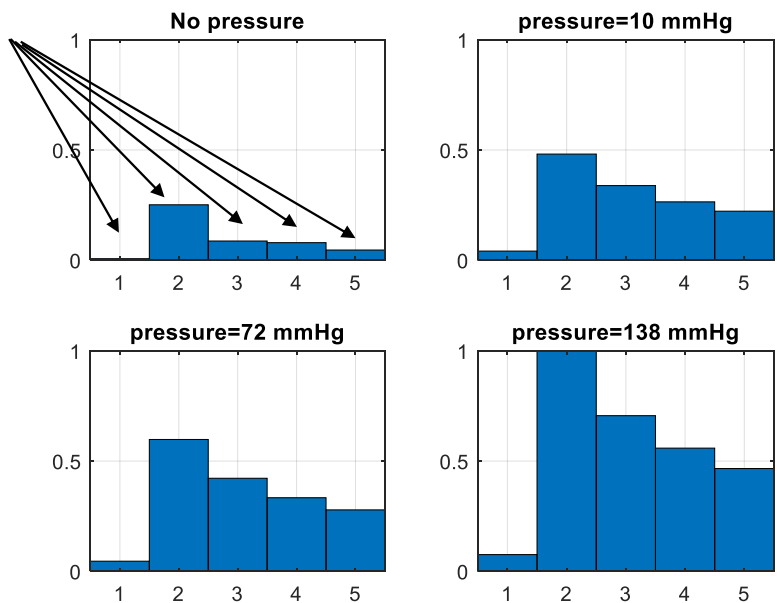

**Fig. (8).** Capacitive sensors power during inflation phase.

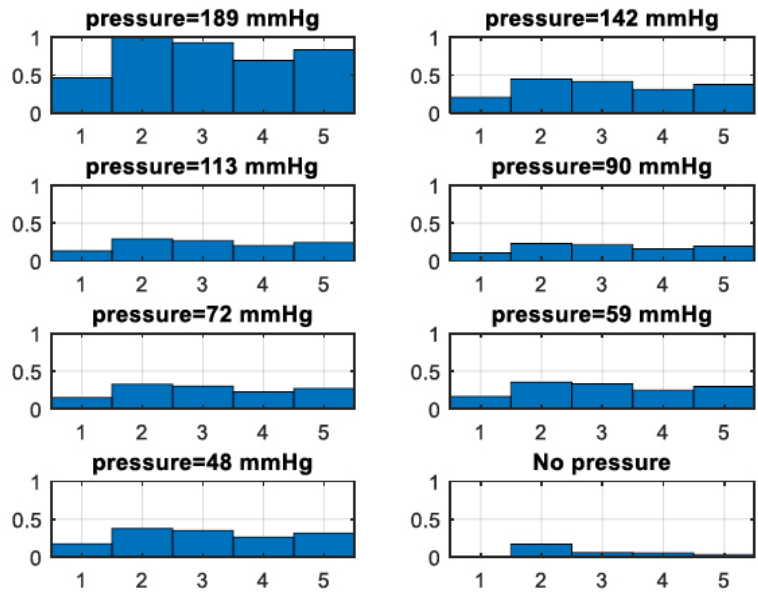

**Fig. (9).** Capacitive sensors power during deflation phase.

$$S_{C_i,y}(P) = \int_{-\infty}^{+\infty} R_{C_i,y}(\tau, P) e^{-i2\pi f\tau}\, d\tau \quad (3)$$

The reproducibility and reliability of a measuring device are an important issue, and statistical tests need to be performed to confirm the accuracy of the data obtained from the pulse measuring device. By comparing the data obtained from the customized pulse meter with that of a calibrated pulse oximeter reference device, some aspects of the performance of the first version of the pulse-taking device were validated, and it showed to be a suitable measuring device [17].

To evaluate the validity of temperament measurement using thermal images, the results of this automatic classification were compared with the results of a standardized questionnaire (for PM) [13], and its accuracy, specificity, and sensitivity were calculated [14, 15]. The accuracy, sensitivity, and specificity of the algorithm on the "test" data have been calculated according to Eqs. (**4**) to (**6**).

$$\text{Accuracy} = \frac{TP+TN}{TP+TN+FP+FN} \quad (4)$$

$$\text{Sensitivity} = \frac{TP}{TP+FN} \quad (5)$$

$$\text{Specificity} = \frac{TN}{FP+TN} \quad (6)$$

Where, TP, TN, FP and FN are true positive, true negative, false positive and false negative, respectively. The results are shown in Table **3**.

Since the correlation coefficient offers a measure of the mutual dependence between two variables x and y, we calculated the Pearson correlation coefficient for assessment of the validity of the extracted pulse features, as shown in Fig. (**3**):

$$r_{x,y} = \frac{Covariance(x,y)}{\delta_x \delta_y} = \frac{\sum_{i=1}^{n}(x_i-\bar{x})(y_i-\bar{y})}{\sqrt{\sum_{i=1}^{n}(x_i-\bar{x})^2}\sqrt{\sum_{i=1}^{n}(y_i-\bar{y})^2}} \quad (7)$$

Where, x, y are feature vectors (extracted from pulse-taking sensors) and represent PM expert wrist pulse characteristics assessment for all patients, respectively.

The above results showed a proper independency of the system function from the PM clinician (in the local clinic, Fig. **2**); thus, the system could be used in a telecare system. Therefore, in order to utilize a telecare system, the capacitive and optical signals and thermal images could be sent to a remote computer (PM expert computer) via the internet for the final diagnosis. A sample page on the PM expert computer is shown in Fig. (**10**).

**Table 3. Performance of different temperaments classification [14, 15].**

| Temperaments | Warm/cold* | Dry/wet** |
|---|---|---|
| Number of Training set* | 27 | 29 |
| Number of Test set | 5 | 5 |
| Accuracy (%) | 94.7 | 74.7 |
| Sensitivity (%) | 95.7 | 66.7 |
| Specificity (%) | 98.2 | 77.8 |
| * Number of participants: 32 | | |
| ** Number of participants: 34 | | |

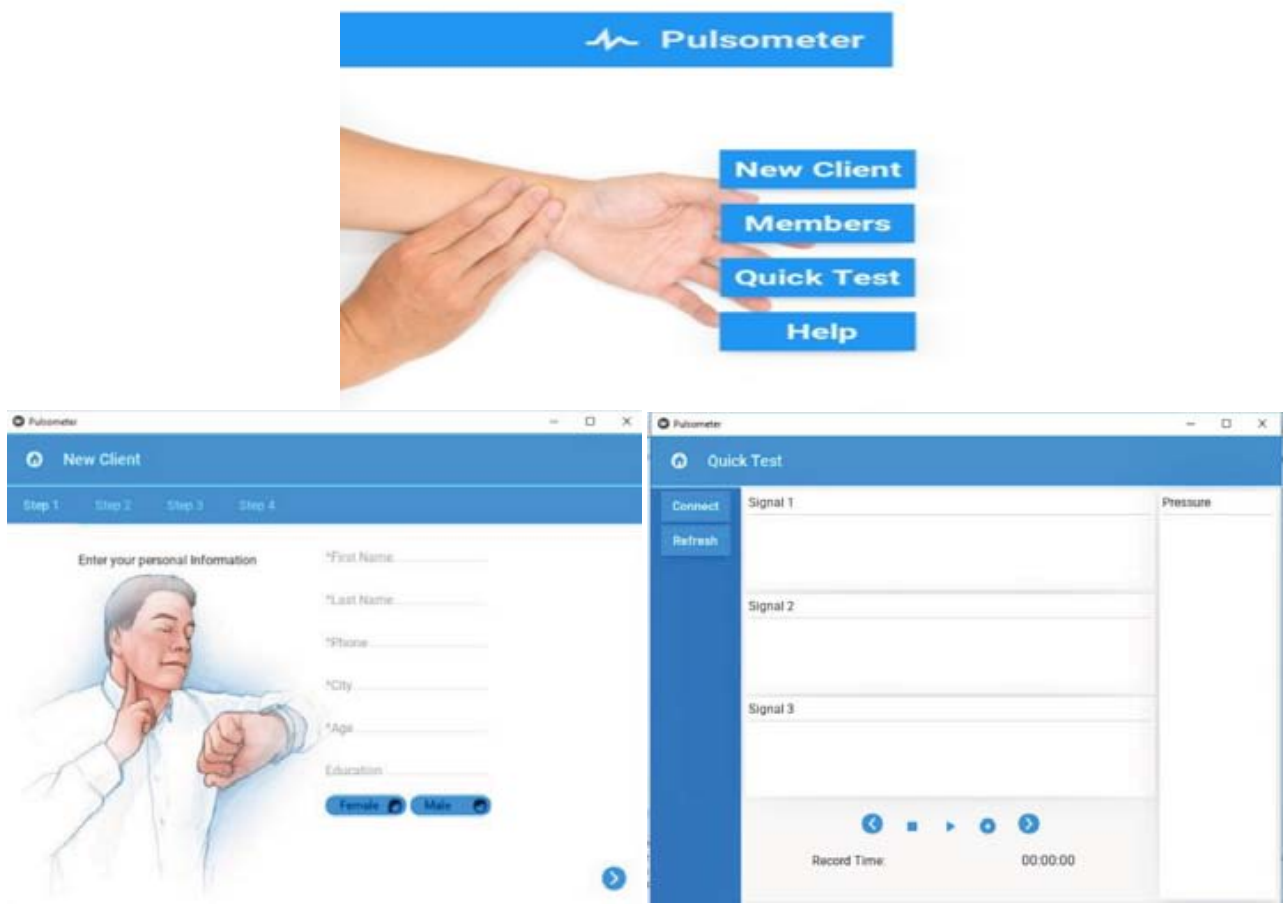

**Fig. (10).** Some user-interface pages of the Telecare system.

## 4. DISCUSSION

The most important means to diagnose diseases in traditional Eastern medicine, and especially in Iran, is by measuring the temperature, humidity, and pulse of the wrist. Generating reproducible results is the main concern in the widespread use of complementary medicine such as PM so that the diagnostic and therapeutic conclusions do not depend on the physician's interpretation of the wrist pulse criteria.

This research can be considered as the first step to the possible use of thermal imaging and wrist pulse measurement obtained with a customized device for assessing temperament in PM. In this study, we combined three measuring instruments to be able to perform the measurements required by PM:

1. Thermal images of body parts

2. Pulse signals at several points on the wrist

3. Blood volume variation in the finger (finger photoplethysmography)

Thermal imaging reveals the distribution of heat on the surface of an object. Recording the temperature distribution patterns in tissue can lead to its physiological interpretation. For example, thermal imaging has been used for the early detection of breast tumors [18, 19]. Another area of the body that has been widely imaged due to its availability is the hand, which has provided useful information for the diagnosis of hand disorders [20]. Thermal images have also been widely used to monitor disease conditions and treatment progress. Indriðadóttir (2019) showed that thermal imaging is a valid tool for quantifying physiological processes such as the skin's response to heat shocks and can be used in such assessments [21]. Also, it has been used for arterial malformations, joint inflammation, bladder stones, gastrointestinal disorders, headaches and migraines, and other diseases that potentially affect the body's temperature patterns [22]. Since temperature and humidity of *Malmas* are some of the determinants of temperament, one part of the present study was to examine the effectiveness of the thermal images obtained from thermal cameras.

Today, the pulse signal is widely used in modern medicine for the early diagnosis of diseases. In PM, the following indicators of the wrist pulse are observed to diagnose patient's condition and diseases [11]:

- Pulse dimensions (Length, Width, Height)
- Pulse velocity

- Pulse frequency
- Pulse strength
- Pulse regularity
- Vascular elasticity

According to these characteristics, traditional medicine physicians have defined different pulse types that can help them to identify a person's temperament along with other above-mentioned factors. At present, traditional medicine practitioners in Iran measure and analyze the wrist pulse with their hands without using a device. Therefore, it is clear that the measurement and diagnosis depend on the physician, and its reproducibility and remote diagnosis can be impaired.

All types of signal/image processing have been done on the remote computer. To compensate for the low sample size and improve the validity and accuracy of the results, a type of cross-validation method was used. This method has been widely used in classification issues when there are not enough experimental samples [23]. In K-fold cross-validation, the original sample is randomly partitioned into K equal-sized subsamples where a single subsample is considered as the test data and the remaining K−1 subsamples are used as training data. The cross-validation process is then repeated K times, with each of the K subsamples used exactly once as the test data. This issue has already been discussed earlier [14, 15], showing that the prediction error of the algorithm increases if the percentage of the training data is too high or too low. Therefore, in order to create a more reliable decision-making system based on this algorithm, future studies must include a greater number of participants.

Despite its long history, PM has been less widely employed than TCM (Traditional Chinese Medicine). It has been emphasized that PM must prove its effectiveness by using valid scientific evidence and reliable devices [24]. Chinese medicine has been equipped with modern devices and equipment for many years. In TCM, the importance of using standard devices to make reliable and reproducible diagnostic results independent of the physician's skills has been well accepted. It has been emphasized that such devices can be effective in understanding the theory of traditional Chinese medicine and its development [25 - 27]. Due to the lack of similar studies specific to PM and due to the similarities between Persian and Chinese traditional medicine, here we mainly review some activities carried out in the field of pulse signals in Chinese medicine.

Chinese experts believe that the radial pulse changes felt in different parts of the wrist are related to the disease of a specific organ. They identified three points on the main artery of the wrist that could be measured by three fingers. For this purpose, a device that simultaneously records the pulse signals of three wrist points was made in 2012 and tested in the laboratory [28]. It has been shown that useful information about the body's function can be gained by examining how this pulse wave energy is distributed anywhere in the artery. Therefore, instead of using one sensor for each point, it is better to consider an array of sensors for each point [29]. Several studies have used the concepts of TCM to distinguish patients with various diseases from healthy people using the wrist pulse signals [30, 31].

Also, pulse wave analysis of the finger (photoplethysmography or PPG) provides much information about the internal organs of the body. Some applications of PPG analysis that have been explored in a number of studies are as follows:

1- Extraction of respiration rate [32]

2- Measurement of blood pressure [33]

3- Determining the vascular stiffness [34]

4- Estimation of cardiac output [35]

5- Separation of healthy people and people with mental disorders [36]

6- Prediction of hypotension during hemodialysis [37].

On the other hand, because the distribution of heat is affected by physiological activity, it can be expected to be related to different temperaments. A part of this research aimed to determine the feasibility of thermal imaging to measure different temperaments in individuals and its compliance with a standardized temperament questionnaire in PM. Since thermal imaging has not been used in PM to measure temperament, it has not been possible to compare the results of this study with similar researches.

It can be assumed that the average temperature and its fluctuations in each body region are the result of the blood flow and metabolism in that area. Also, the heat distribution at the surface of an area depends on the characteristics and thickness of the tissue and the vascular structure under that area [38]. Some studies have confirmed the use of thermal imaging for diagnostic evaluation in traditional medicine, which is in line with the findings of the present study on the use of thermal imaging for assessing the qualities of warm/cold temperament in traditional medicine. Infrared thermography is widely used in studies of meridians in acupuncture, as well as in the diagnosis and treatment of diseases in Chinese medicine [39 - 42].

Thermal imaging has an advantage over using a common thermometer. Imaging can determine the temperature and its variations in the entire region, while a thermometer only determines the temperature at the point of contact. Numerous studies have shown that mental and emotional states affect the automatic control of facial skin temperature, which can be recorded by thermal imaging of the face. These studies have been reviewed by Cardone [43].

## CONCLUSION

This study showed that designing and implementing a customized system for recording and quantifying temperature (by a relatively inexpensive camera), humidity, and pulse wave in PM is possible and can reduce the dependence on the interpretation of the wrist pulse and patient's condition by the physician. This system makes it possible to make traditional medicine diagnoses remotely, which is an effective solution in widespread pandemics.

## ETHICS APPROVAL AND CONSENT TO PARTI-

## CIPATE

This study was approved by the Medical Ethics Committee.

## HUMAN AND ANIMAL RIGHTS

Not applicable.

## CONSENT FOR PUBLICATION

Not applicable.

## FUNDING

None.

## CONFLICT OF INTEREST

The authors declare no conflict of interest, financial or otherwise.

## ACKNOWLEDGEMENTS

The authors wish to thank the Infrared Technologists Company, especially Mr. Alidoosti, for providing the thermal camera. Also, the authors are grateful to Ms. Mohadeseh Haydarpanah for her part in collecting and categorizing the data.